\documentclass[10pt,conference]{IEEEtran}
\ifCLASSINFOpdf
\else
\fi
\usepackage[utf8]{inputenc}
\usepackage{amsmath,amssymb}
\usepackage{tikz}
\usepackage{algpseudocode}
\usepackage{listings}
\usepackage[linesnumbered,ruled,vlined]{algorithm2e}
\usepackage{xcolor}
\usepackage{multirow}
\usepackage{svg}
\usepackage{newtxtext}
\usepackage{booktabs}
\usepackage{amsthm}
\usepackage{url}

\begin{document}
\title{DeepTx: Real-Time Transaction Risk Analysis via Multi-Modal Features and LLM Reasoning}

\author{\IEEEauthorblockN{Yixuan Liu}
	\IEEEauthorblockA{
    Nanyang Technological University\\
    Singapore\\
        liuy0255@e.ntu.edu.sg}
	\and
        \IEEEauthorblockN{Xinlei Li}
	\IEEEauthorblockA{
    Nanyang Technological University\\
    Singapore\\
        xinlei003@e.ntu.edu.sg}
	\and
	\IEEEauthorblockN{Yi Li}
	\IEEEauthorblockA{
    Nanyang Technological University\\
    Singapore\\
        yi\_li@ntu.edu.sg}
    }

\maketitle

\begin{abstract}
Phishing attacks in Web3 ecosystems are increasingly sophisticated, exploiting deceptive contract logic, malicious frontend scripts, and token approval patterns. We present DeepTx, a real-time transaction analysis system that detects such threats before user confirmation. DeepTx simulates pending transactions, extracts behavior, context, and UI features, and uses multiple large language models (LLMs) to reason about transaction intent. A consensus mechanism with self-reflection ensures robust and explainable decisions. Evaluated on our phishing dataset, DeepTx achieves high precision and recall (demo video: \url{https://youtu.be/4OfK9KCEXUM}). 
\end{abstract}

\begin{IEEEkeywords}
Blockchain security, Phishing detection, Transaction semantics
\end{IEEEkeywords}

\IEEEpeerreviewmaketitle
\section{Introduction}

Blockchain users frequently interact with decentralized applications (DApps) through Web3 wallets
and browser-based frontends. These interactions often require users to sign transactions that are
encoded in low-level calldata, which hides critical semantic information such as the actual
destination address, function being called, or asset being transferred. As a result, attackers
increasingly exploit this semantic gap by crafting phishing transactions that mislead users through
benign-looking interfaces but invoke malicious behaviors after signing.

A recent example of such an attack occurred in February 2025, when Bybit, one of the largest
cryptocurrency exchange in the world, suffered a loss of over 400{,}000 ETH (around \$1.5 billion)
due to a manipulated wallet interface~\cite{bybit2025}. The attacker injected a forged User
Interface (UI) that displayed a legitimate transaction to the operator, while the signed calldata
secretly executed a privileged operation such as transferring ownership to a malicious contract.
This enabled subsequent unauthorized asset withdrawals. The attack bypassed all traditional
on-chain protections and signature verifications, highlighting a critical security risk arising not
from contract vulnerabilities, but from misleading user interactions and transaction
representations.

% Existing tools such as block explorers, static analyzers, and reputation systems focus primarily on known threats, label-based filtering, or post-hoc detection. These approaches offer limited defense against novel phishing schemes, interface forgery, or transaction obfuscation at signing time. To address this gap, we introduce \textbf{DeepTx}, a real-time semantic analysis system for detecting phishing and deceptive transactions before user confirmation.

Several tools have been developed to mitigate phishing transactions, each adopting different
detection strategies with varying levels of openness and effectiveness. PTXPhish~\cite{chen2025dissecting} is an open-source tool that analyzes behavioral and contextual features within
transactions. However, its rule-based design lacks adaptability and often fails to detect
previously unseen attack vectors.
In contrast, ScamSniffer~\cite{scamsniffer} and Pocket Universe~\cite{pocketuniverse} focus on call
data simulation, matching results against hard-coded patterns.
While ScamSniffer partially discloses its block lists and browser extension code, both tools retain
proprietary detection logic, limiting their transparency and extensibility.
Forta~\cite{forta}, a decentralized alerting system, employs community-operated bots to monitor
live on-chain activity.
Although Forta supports real-time data streams, its alerts are typically issued only after
transactions have been mined, and together with other tools that rely on static rules or post-hoc
pattern recognition, it offers limited protection against phishing tactics involving UI forgery,
calldata obfuscation, or execution flows at the time of signing. An early LLM-based attempt at transaction analysis exists, but it lacks a clear architecture and an end-to-end implementation~\cite{pydeeptx_2025}.

In this paper, we propose DeepTx, a tool that simulates user-intended transactions as if they were
already signed, and extracts multi-modal features including behavioral traces, contextual
behaviors, and user interface indicators.
It then uses LLMs to analyze the transaction's intent and potential security risks.
The system generates a natural language risk report that includes a severity level, explanatory
description, and suggested user action.
To enhance reliability, DeepTx performs consensus checking across multiple LLMs and applies
self-reflection techniques to resolve ambiguous results.

We summarize our contributions as follows:
\begin{itemize}
  \item We propose \textbf{DeepTx}, a real-time transaction semantics analysis tool that detects
  phishing and deceptive behaviors before user confirmation.
  \item We design a multi-modal feature extraction pipeline, incorporating behavioral traces,
  contextual signals, and UI metadata from simulated user transactions.
  \item We integrate large language models with a consensus checking and self-reflection mechanism
  to produce reliable and explainable security assessments.
  \item We publicly release both the tool and the accompanying dataset, which captures the full
  phishing lifecycle---including frontend interaction scripts, confirmed victim transactions, and
  detailed call chain data. All resources are available at \url{https://github.com/yxsec/DeepTx}.
\end{itemize}

\section{System Design and Implementation}

\begin{figure}[h]
    \includegraphics[width=1\linewidth]{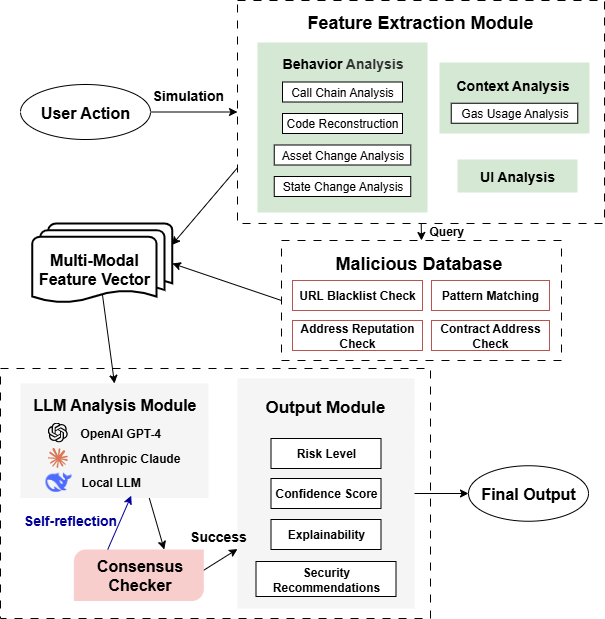}
    \caption{Overview of DeepTx}
    \label{fig:system-architecture}
\end{figure}

DeepTx performs pre-signing transaction simulation and multi-perspective analysis to detect phishing and deceptive behaviors. It extracts features from transaction execution, evaluates contextual and UI signals, queries known malicious indicators, and synthesizes a final risk assessment using a consensus-guided LLM reasoning module. Figure~\ref{fig:system-architecture} outlines the overall architecture.

\subsection{System Overview}
When a user initiates a transaction, DeepTx simulates its execution in a forked EVM state.
It extracts semantic features from the trace, including behavioral effects (e.g., call chain, token
flows, and state changes), contextual signals (e.g., gas usage), and optional UI features (e.g.,
JavaScript constructing the transaction).
These features are augmented by querying a malicious database with blacklists and pattern rules.

The extracted information is encoded into a multi-modal feature vector, partitioned into four
modules \texttt{(behavior, context, UI, database)}, each assigned a custom weight summing to one.
This vector is submitted to $n$ LLMs, each producing a structured response containing the
transaction's risk level, confidence, explanation, and user recommendations.

A consensus mechanism ensures better robustness.
If the LLMs agree, one model summarizes the results into the final decision.
If disagreement occurs, DeepTx enters a self-reflection phase where each model is re-prompted with
others' responses as counterexamples.
If no agreement is reached after a fixed number of iterations, DeepTx falls back to weighted voting
based on confidence scores.
The final output is composed from the last round's outputs and shown to the user.

\subsection{Transaction Simulation}
DeepTx uses local or remote simulation to replay transactions.
It supports Foundry’s \texttt{cast run}~\cite{foundry} with Anvil and Tenderly's simulation
API~\cite{tenderly}.
These provide execution traces, calldata, internal calls, and storage changes, enabling downstream
semantic analysis.
No state is broadcast on-chain, ensuring safety.

\subsection{Feature Extraction Module}
After simulating the transaction, DeepTx extracts a structured set of semantic features to
characterize its intent and effect.
The extracted features are grouped into three categories: behavioral, contextual, and UI-related.
These collectively form a multi-modal representation that supports subsequent LLM reasoning.

\subsubsection{Behavioral Features}
Behavioral analysis captures the concrete effects of the transaction on the EVM state. DeepTx extracts the full call chain from the execution trace, including all internal and external function invocations, along with the corresponding code segments executed. For each contract involved in the call chain, the system retrieves its code: if the contract is verified, DeepTx fetches the source-level functions using the Etherscan API~\cite{etherscan}; otherwise, it uses Heimdall~\cite{heimdall} to decompile the bytecode and approximate its logic.

Asset-related actions are also identified.
DeepTx tracks native ETH and ERC-20 token transfers by analyzing value fields and standard
interfaces, such as \texttt{transfer} and \texttt{transferFrom}.
Additionally, storage writes (\texttt{SSTORE}) are collected and analyzed to detect role updates,
ownership changes, and balance modifications.
These behaviors offer insight into the transaction's operational intent.

\subsubsection{Contextual Features}
Contextual features capture the transaction environment and help detect anomalous or deceptive behavior. DeepTx compares the gas limit and actual usage, flagging transactions with excessive unused gas as potentially misleading. It also analyzes the effective gas price relative to the base fee to detect possible transaction acceleration or frontrunning. Sender address and nonce patterns are used to identify rapid transaction sequences, which may result from automation or repeated confirmations triggered by phishing interfaces.

\subsubsection{UI Features}
If the transaction is initiated from a DApp frontend, DeepTx optionally performs interface-level analysis. It statically analyzes the associated JavaScript code to identify logic responsible for calldata construction and signature initiation, including inline and external scripts when available.

To support phishing detection, DeepTx also extracts the interaction page’s URL and main domain. These signals are combined with threat intelligence data and encoded into the feature vector. 
UI analysis is performed only when such metadata is accessible at the time of simulation.

\subsection{Malicious Database}
DeepTx queries multiple malicious data sources to enrich semantic reasoning.
These include: (1) ScamSniffer-based domain and address blacklists, (2) contract address tags, and
(3) manually defined calldata or function selector patterns indicative of scams.
These results are incorporated into the LLM reasoning module.

\begin{figure}
    \includegraphics[width=1\linewidth]{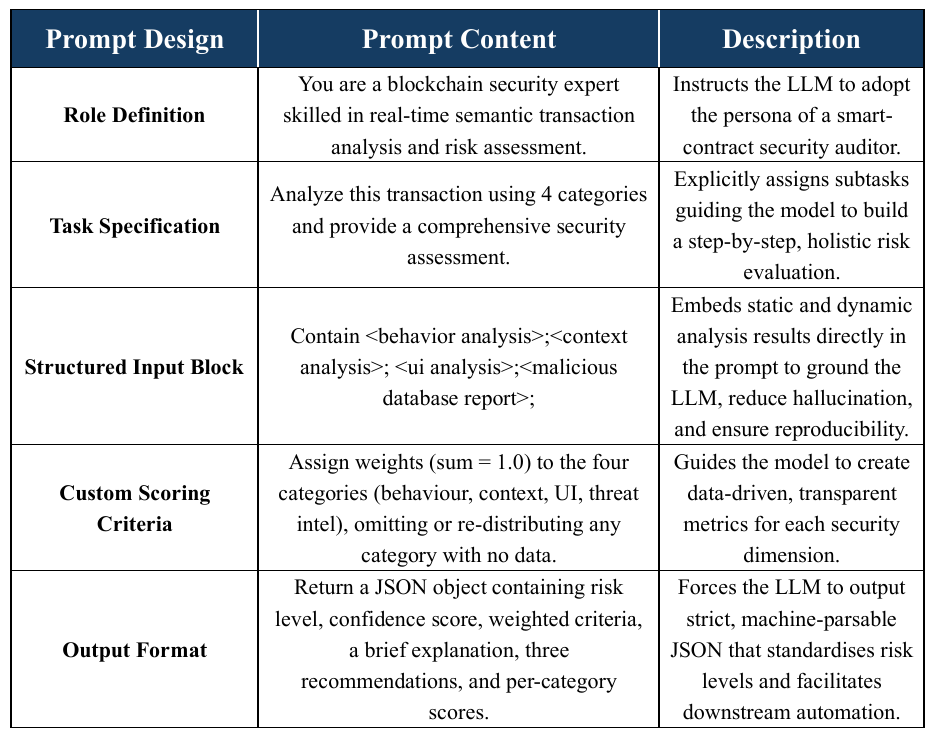}
    \caption{LLM prompt template used in DeepTx (generalized for $n$ models)}
    \label{fig:prompt-template}
\end{figure}

\begin{algorithm}[t]
\scriptsize
\DontPrintSemicolon
\KwIn{Initial LLM outputs $\{O_1, O_2, \dots, O_n\}$}
\KwOut{Final consensus decision $O^*$}
Initialize $R \gets 0$ (round counter), $M \gets 3$ (default max rounds)\;

\While{$R < M$}{
    Extract predicted risks $\{r_i\}_{i=1}^n$ and confidences $\{c_i\}_{i=1}^n$ from $\{O_i\}$\;

    \If{$r_1 = r_2 = \dots = r_n$}{
        $O^* \gets$ summarize($O_1, O_2, \dots, O_n$) via a primary LLM\;
        \Return $O^*$
    }

    \tcp{Perform self-reflection with counterexamples}
    \For{$i \in \{1,\dots,n\}$}{
        Let $O_{\text{own}} \gets O_i$, $O_{\text{counter}} \gets \{O_j \mid j \ne i\}$\;
        $O_i' \gets$ selfReflect($O_{\text{own}}, O_{\text{counter}}$)\;
    }

    Update all $O_i \gets O_i'$ for $i \in \{1,\dots,n\}$\;
    $R \gets R + 1$\;
}

\tcp{Fallback: weighted voting by confidence}
Initialize score map $S[r] \gets 0$ for all possible labels $r$\;
\For{$i \in \{1,\dots,n\}$}{
    $S[r_i] \gets S[r_i] + c_i$\;
}
$r^* \gets \arg\max_r S[r]$\;
\Return any $O_i$ such that $r_i = r^*$\;
\caption{Consensus Checker with Self-Reflection and Weighted Voting (generalized to $n$ models)}
\label{alg:consensus}
\end{algorithm}

\subsection{Semantic Reasoning with LLMs}
The final stage is to classify whether the transaction is phishing or malicious, estimate its risk level and intent, and provide a human-interpretable explanation before user confirmation.

\subsubsection{Prompt Design}
To support reliable and explainable reasoning, DeepTx constructs a structured prompt that encodes
all available information, including the behavior trace, gas context, transaction-specific code
snippets, frontend scripts, and threat intelligence from the malicious database.
The prompt assigns the LLM the role of a blockchain security analyst and specifies an evaluation
task across four categories: behavior, context, UI, and database indicators.

As illustrated in Figure~\ref{fig:prompt-template}, the prompt includes a strict output schema,
requiring each model to return a JSON object containing the predicted risk label (\texttt{safe},
\texttt{suspicious}, or \texttt{malicious}), a numerical confidence score, a justification for the
decision, a transaction summary, feature importance weights summed to 1, and a list of actionable
security recommendations.
This standardization ensures consistency across different models and supports consensus checking.

\subsubsection{Multi-Model Reasoning and Consensus}
The structured prompt is submitted to $n$ LLMs—such as GPT-4, Claude, local fine-tuned models, or
other security-focused systems—which independently generate risk assessments.
If all models produce consistent risk labels, one LLM is selected to summarize the results into a
final user-facing report.
Otherwise, DeepTx initiates a self-reflection phase, as outlined in Algorithm~\ref{alg:consensus}.
In this phase, each model is re-prompted with the outputs of the others provided as counterarguments
and asked to reassess its original decision.
This iterative reasoning process continues until either a consensus is achieved or a predefined
round limit $M$ is reached.

\subsubsection{Fallback Voting Mechanism}
If no consensus is reached after $M$ rounds of self-reflection, DeepTx falls back to a weighted
voting strategy, as shown in the second part of Algorithm~\ref{alg:consensus}. The final risk label
is selected based on the cumulative confidence-weighted votes for each class.
Specifically, the total score for each candidate label $r$ is computed as:
{\(\text{Score}(r) = \sum_{i=1}^{n} \mathbb{I}[r_i = r] \cdot c_i\)}
where $r_i$ is the label assigned by model $i$ with associated confidence $c_i$, and $\mathbb{I}[\cdot]$ is the indicator function. The final predicted label $r^*$ is selected as:
{
\(
r^* = \arg\max_{r} \text{Score}(r).
\)
}
The output from any model satisfying $r_i = r^*$ is used as the final decision.
The system merges all available outputs from the final round to produce a user-facing summary that
includes supporting explanations, risk rationale, and suggested actions.

\begin{figure}[t]
    \centering
    \includegraphics[width=1\linewidth]{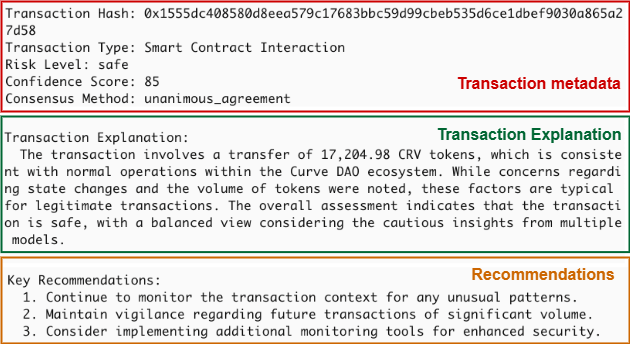}
    \caption{Final analysis summary produced by DeepTx. The output provides a human-readable explanation, transaction-specific confidence score, and actionable recommendations for user decision making.}
    \label{fig:final-analysis-summary}
\end{figure}

\subsection{Output Module}
While the user-facing summary shown in Figure~\ref{fig:final-analysis-summary} provides a concise
view of the transaction's risk level, explanation, and recommendations, DeepTx also generates a
corresponding structured JSON report that contains additional information not visible in the
summary view.
Specifically, the JSON file includes detailed component-level scores and reasoning across four
semantic dimensions---behavior patterns, gas context, UI indicators, and threat intelligence---each
with an assigned weight reflecting its contribution to the final decision.
It also records metadata from the consensus process, such as whether the decision was reached
unanimously or through weighted voting, the number of self-reflection rounds used, and the identity
of the primary model that produced the final output.
This structured format supports logging, auditing, and integration into automated security
workflows.

\subsection{Tool Usage}
After configuring the \texttt{.env} file, DeepTx provides two modes—Historical Transaction and Simulation—to analyze existing on-chain transactions or execute a specified contract function for simulated execution, returning a structured semantic report in both cases.
\section{Preliminary Evaluation}

To provide a preliminary evaluation of DeepTx, we constructed a dataset of phishing victim transactions with associated web data and conducted ablation experiments under different UI feature settings.

\subsection{Dataset Collection}

We manually labeled 12 phishing cases from two sources, including 5 simulation-based phishing
challenges from UnPhishable~\cite{UnPhishable} and 7 real-world cases collected from ScamSniffer
database and archived phishing websites~\cite{wayback}.
These cases cover a range of semantic and deception strategies, including malicious token
approvals, proxy spoofing, and impersonation via fake interfaces.
For each transaction, we collected (1) the associated frontend webpage and JavaScript if available,
(2) verified or decompiled smart contract code, and (3) the original victim
transaction from the chain.

To enable false positive evaluation, we included 2 benign transactions selected from verified
protocols (e.g., Aave~\cite{aave}) and manually confirmed them to be non-malicious.

\subsection{Experiment Setup and Metrics}

All transactions were replayed in a forked EVM state to reproduce their original blockchain
execution environment. DeepTx then analyzed each transaction through a sequence of analysis stages,
including behavioral simulation, contextual feature extraction, UI signal analysis, and LLM-based
reasoning.

We evaluated three LLM configurations within DeepTx: \texttt{gpt-4o-mini}, \texttt{gpt-3.5-turbo},
and \texttt{gpt-4o}, all accessed via the OpenAI API.
The reported results use \texttt{gpt-4o} as the default model.

Each configuration was executed over three independent runs to account for non-determinism in LLM
output.
A phishing transaction is considered successfully detected if it is labeled as either
\texttt{malicious} or \texttt{suspicious}.
A false positive is recorded when a benign transaction is incorrectly flagged as risky.
We report the mean and standard deviation of precision, recall, and F1 score across runs.

\begin{table}[h]
\centering
\caption{Detection performance on phishing and benign cases (mean ± std)}
\label{tab:overall-performance}
\begin{tabular}{lccc}
\toprule
\textbf{Configuration} & \textbf{Precision} & \textbf{Recall} & \textbf{F1 Score} \\
\midrule
Full DeepTx & \textbf{1.00 ± 0.00} & \textbf{0.89 ± 0.05} & \textbf{0.94 ± 0.03} \\
No UI \& No Database       & 0.92 ± 0.14 & 0.22 ± 0.05 & 0.35 ± 0.06 \\
\bottomrule
\end{tabular}
\end{table}

As shown in Table~\ref{tab:overall-performance}, DeepTx achieves high precision and recall under
the full configuration.
Disabling UI features significantly reduces recall, highlighting the importance of interface-level
signals.

\section{Conclusion}
DeepTx is a real-time transaction analysis system that inspects user-intended EVM transactions
before signing.
By simulating execution and extracting behavioral, contextual, and interface-level features, the
system uses multiple LLMs with consensus and self-reflection to assess transaction intent and
potential risks.
Evaluation on our phishing dataset demonstrates that DeepTx achieves high precision and recall,
offering practical protection against deceptive transactions.
Future work will extend the dataset to cover more scenarios and incorporate blind signature
analysis to further improve risk assessment.

\section{Acknowledgements}
This research is supported by the Singapore Ministry of
Education Academic Research Fund Tier 2 (T2EP20224-0003) and the Nanyang Technological University Centre
for Computational Technologies in Finance (NTU-CCTF).
Any opinions, findings, and conclusions or recommendations
expressed in this material are those of the author(s) and do not
necessarily reflect the views of MOE and NTU-CCTF.

\bibliographystyle{IEEEtran}
\bibliography{reference}

\end{document}